\documentclass[runningheads]{llncs}
\usepackage[T1]{fontenc}
\usepackage{graphicx}
\usepackage{hyperref}
\usepackage{color}

\usepackage{amssymb}
\usepackage{subfigure}
\usepackage[acronym]{glossaries}
\usepackage{subcaption}
\usepackage[misc,geometry]{ifsym}
\usepackage{orcidlink}

\newacronym{iac}{IAC}{intracranial arterial calcification}
\newacronym{ct}{CT}{computer tomography}
\newacronym{ssim}{SSIM}{structural similarity index measure}
\newacronym{ist3}{IST-3}{third International Stroke Trial}
\newacronym{roi}{ROI}{region of interest}
\newacronym{mri}{MRI}{magnetic resonance imaging}

\begin{document}
\title{Pre-processing and quality control of large clinical CT head datasets for intracranial arterial calcification segmentation}
\titlerunning{Pre-processing and quality control of large clinical CT head datasets}
%
\author{
Benjamin Jin\inst{1}\textsuperscript{(\Letter)}\orcidlink{0009-0006-2710-7248} \and
Maria del C. {Valdés Hernández}\inst{1,2}\orcidlink{0000-0003-2771-6546} \and
Alessandro Fontanella\inst{3}\orcidlink{0000-0003-4885-9407} \and
Wenwen Li\inst{1,3}\orcidlink{0000-0003-3710-1967} \and
Eleanor Platt\inst{3} \and
Paul Armitage\inst{4}\orcidlink{0000-0003-2790-5478} \and
Amos Storkey\inst{3}
Joanna M. Wardlaw\inst{1,2}\orcidlink{0000-0002-9812-6642} \and
Grant Mair\inst{1,5}\orcidlink{0000-0003-2189-443X}
}
\authorrunning{B. Jin et al.}
%
\institute{
Centre for Clinical Brain Sciences, University of Edinburgh,  Edinburgh,UK\\
    \email{b.jin@ed.ac.uk}\and
UK Dementia Research Institute, University of Edinburgh, Edinburgh, UK \and
School of Informatics, University of Edinburgh, Edinburgh, UK \and
Department of Infection, Immunity and Cardiovascular Disease, University of Sheffield, Sheffield, UK \and
Neuroradiology, Department of Clinical Neurosciences, NHS Lothian, Edinburgh, UK
}
\maketitle              
\begin{abstract}
As a potential non-invasive biomarker for ischaemic stroke, \acrfull{iac} could be used for stroke risk assessment on CT head scans routinely acquired for other reasons (e.g. trauma, confusion). Artificial intelligence methods can support \acrshort{iac} scoring, but they have not yet been developed for clinical imaging. Large heterogeneous clinical CT datasets are necessary for the training of such methods, but they exhibit expected and unexpected data anomalies. Using CTs from a large clinical trial, the third International Stroke Trial (IST-3), we propose a pipeline that uses as input non-enhanced CT scans to output regions of interest capturing selected large intracranial arteries for \acrshort{iac} scoring. Our method uses co-registration with templates. We focus on quality control, using distribution of information along the z-axis of the imaging to group and apply similarity measures triaging assessment of individual image series. Additionally, we propose superimposing thresholded binary masks of the series to inspect large quantities of data in parallel. We identify and exclude unrecoverable samples and registration failures. In total, our pipeline processes 10,659 CT series, rejecting 4,322 (41\%) in the entire process, 1,450 (14\% of the total) during quality control, and outputting 6,337 series. Our pipeline enables effective and efficient region of interest localisation for targeted \acrshort{iac} segmentation.

 \and deep learning
\end{abstract}
\setcounter{footnote}{0} 

\section{Introduction}
Stroke occurs suddenly for patients -- from one moment to another, they are struck by a life-altering event. With evidence for associations between \acrfull{iac} and risk of ischaemic stroke \cite{Li2023,Banerjee2017}, non-invasive imaging biomarkers such as \acrshort{iac} enable preventive approaches to stroke management. The background of \acrshort{iac} as a biomarker for stroke is atherosclerosis, a systemic cardiovascular disease. During disease progression, the lumen of arteries increasingly narrows, leading to decreased blood flow through the affected vessels. This is caused by a build-up of plaque consisting of fatty and fibrous materials, which is often calcified \cite{Libby2019}. The latter is critical, because calcium (due to its high density) is visible on non-enhanced \acrfull{ct} scans. These are routinely acquired, e.g. for patients presenting with trauma or confusion, and thus are primed for early risk assessment of a broad population.

Calcium quantification in the coronary arteries on routine \acrshort{ct} is utilised to assess the risk of future heart attacks dating back to the seminal work by Agatston et al. about three decades ago \cite{Agatston1990}. Since then, it has been established as part of clinical practice \cite{Gupta2022}. \acrshort{iac} could be used similarly in the future to assess the risk of stroke. But what is missing for \acrshort{iac} risk scoring to be adopted in clinical practice? Although associations between \acrshort{iac} and ischaemic stroke have been confirmed \cite{Li2023,Banerjee2017}, longitudinal studies are missing to describe the precise causal relationship necessary to perform risk prediction. Such studies require a large calcium scoring effort. Visual calcium scoring methods can offer rapid manual assessment of calcium burden \cite{Wardlaw2015}. More recently, deep learning methods, which learn to recognise patterns from data \cite{LeCun2015}, were developed on research data to automate quantitative scoring \cite{Bortsova2017,Bortsova2021}. Yet, their performance on clinical data with diverse populations and heterogeneous scans remains elusive.

Here, we describe a pipeline for heterogeneous clinical \acrshort{ct} head scans preparing them for deep learning training to segment \acrshort{iac}. We follow and adapt the approaches by Bortsova et al. in \cite{Bortsova2021} and Fontanella et al. in \cite{Fontanella2023}. The former pre-process their data for a similar task, i.e. \acrshort{iac} segmentation. The latter devise their pipeline for the same data we use here, but for stroke lesion segmentation. Our main contributions are the description and evaluation of the integration of effective quality control into a pipeline processing a large clinical dataset of CT scans, and the outline, definition, and implementation as a digital resource, of the corresponding \acrfullpl{roi} for future \acrshort{iac} segmentation.

\section{Methods}   
To build our pipeline we use data from the \acrshort{ist3} \cite{IST-3_collaborative_group_2012,IST-3_collaborative_group_2015}. The \acrshort{ist3} was a large multi-centre randomised-controlled trial of intravenous alteplase treatment among 3035 patients with acute ischaemic stroke from 156 centres in 12 countries. Due to data corruption and restriction to non-enhanced CT scans, we use a large subset of the data consisting of 10,659 CT series from 2,578 patients. Multiple series can originate from the same scan with differences in soft tissue or bone kernel imaging, patient orientations, separate skull base and skull vault series, and localisers (supplementary Figure 1).

We heavily draw from the work by Fontanella et al. in \cite{Fontanella2023}. Mainly, we convert the series in native DICOM (Digital Imaging and Communications in Medicine) format \cite{dicom} to the NIfTI (Neuroimaging Informatics Technology Initiative) standard \cite{nifti} with dcm2niix \cite{Li2016}, limit our downstream processing to axial series, remove localisers, and co-register the remaining series with two age-appropriate \acrfull{mri} templates \cite{Farrell2009}\cite{Armitage2024} using the FMRIB's Linear Image Registration Tool (FLIRT) for affine registration \cite{Jenkinson2001,Jenkinson2002}. For more details on this part of the pipeline please refer to \cite{Fontanella2023}. We diverge from the previous pipeline \cite{Fontanella2023} by including bone kernel and incomplete CT head series. We are interested in \acrshort{iac} rather than brain tissue, which is visible and better resolved on imaging processed using a bone kernel. The decision entails increased heterogeneity, as bone kernel imaging appears sharper and noisier. However, it will enable the applicability of a deep learning model to a wider range of CT series. Secondly, the main \acrshortpl{roi} for calcium scoring are in the lower part of the head. Incomplete series including these \acrshortpl{roi} are not required to include complete brain tissue, increasing our dataset size compared to Fontanella's work \cite{Fontanella2023}.

We follow the approach in \cite{Bortsova2021} by defining regions around the arteries in the \acrshort{mri} templates and transferring the regions after co-registration into the native space of the original \acrshort{ct} scans. In addition to a region around the cavernous segment of the internal carotid artery and the M1 segment of the middle cerebral artery (anterior Circle of Willis), we define regions for the vertebral arteries from the foramen magnum up to their merger as the basilar artery (posterior Circle of Willis), see supplementary Figure 2 and Figure 3. The \acrshortpl{roi} are available from \cite{Jin2024}. Lastly, we only retain CT series which include at least half of the volume of one \acrshort{roi}. Our additional quality control of the CT series and derived \acrshortpl{roi} is described in detail in the sections below.

\subsection{Slice information presence}
\label{subsec:information_presence}

We propose a simple measure to determine the amount of the patient's head visible per slice to classify groups of \acrshort{ct} series covering different parts of the head. The measure is not to be confused with Shannon entropy \cite{Shannon1948} often used to measure information in data. Entropy measures the spread of the slice's distribution of attenuation values and thus is sensitive to differences in noise or contrast and changes of the attenuation distribution throughout the head. In contrast, information presence exploits the calibration of CT attenuation values to Hounsfield Units (HU). HU represent matter density, so low values denote air (absence of the patient's head). Given a CT series with attenuation values $\mathbf{A} \in \mathbb{R}^{x \times y \times z}$, we define $a_{min}$ as a minimum threshold and with constant tolerance $t$, $0 \leq t \ll 1$ (we experimentally choose $t = 0.05$, commonly corresponding to $a_{min} \approx \text{-}800$):

\begin{equation}
    a_{min} = min(\mathbf{A}) + (max(\mathbf{A}) - min(\mathbf{A})) * t    
\end{equation}

We then define \textit{informative} voxels for the values $\mathbf{A}_s$ of slice $s$ as a set $N_s$:

\begin{equation}
    N_s = \{a \in \mathbf{A}_s | a > a_{min}\}
\end{equation}

Finally, information presence $I_s$ is the slice's proportion of informative voxels:

\begin{equation}
    I_s = \frac{|N_s|}{|\mathbf{A}_s|}
\end{equation}

The information presence shows how much patient imaging is visible in a slice. It also shows how much information is generally available across slices and helps identify inconsistencies in the entire dataset.

\subsection{Grouped structural similarity index measure (SSIM) quality control}
\label{subsec:ssim_quality control}

\begin{figure}
    \centering
    \includegraphics[width=0.24\textwidth]{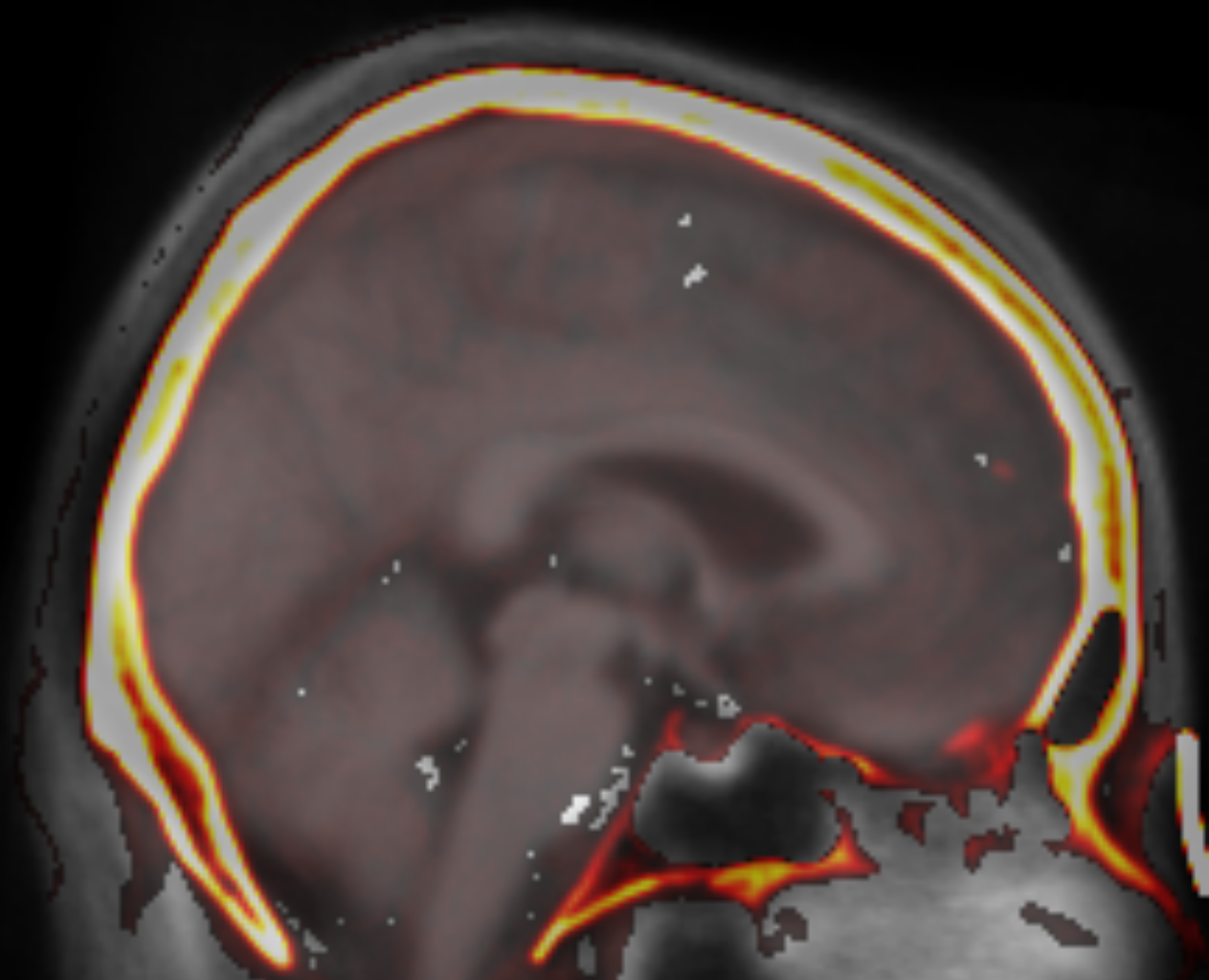}\label{subfigure:complete_group}
    \includegraphics[width=0.24\textwidth]{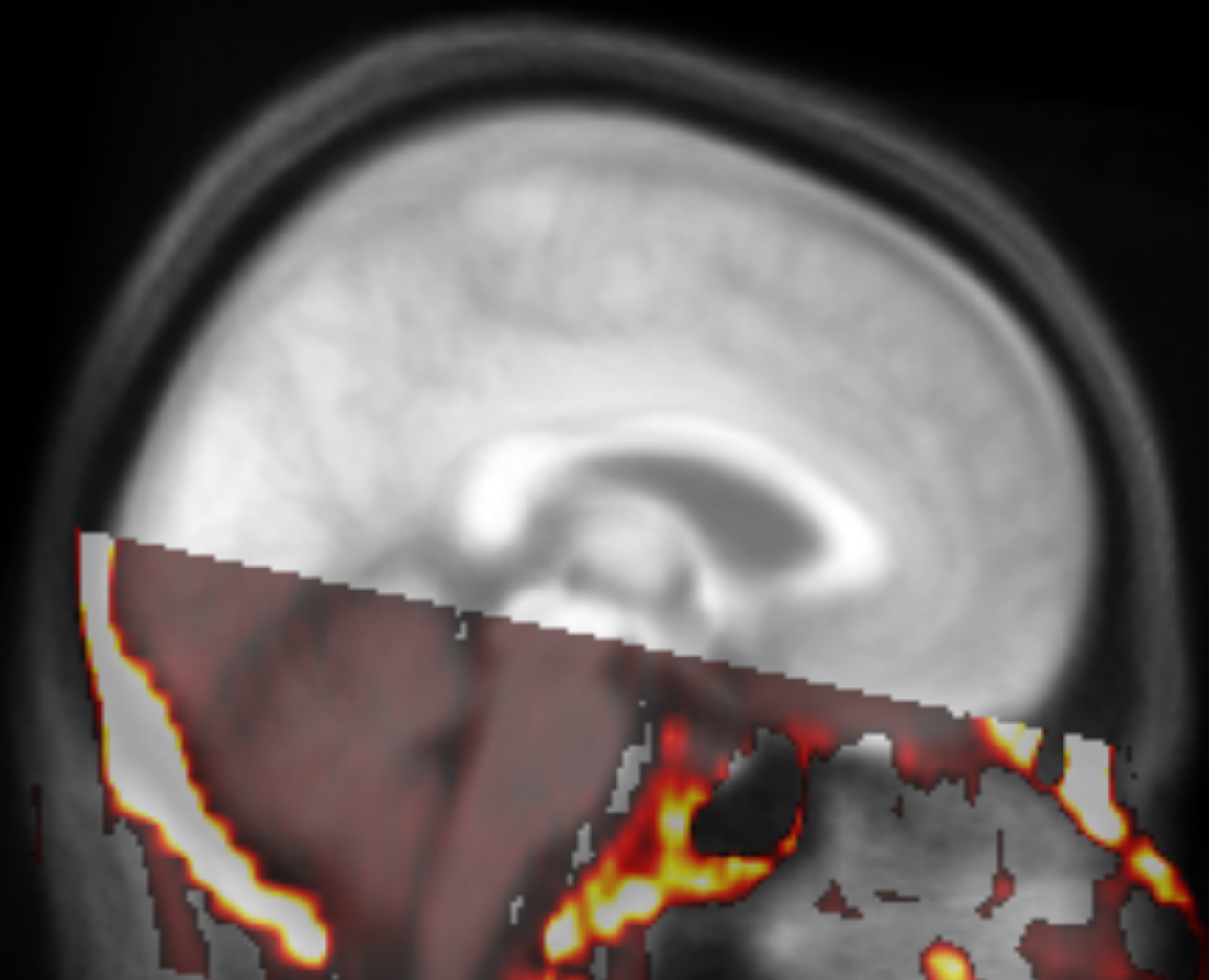}\label{subfigure:skull_base_group}
    \includegraphics[width=0.24\textwidth]{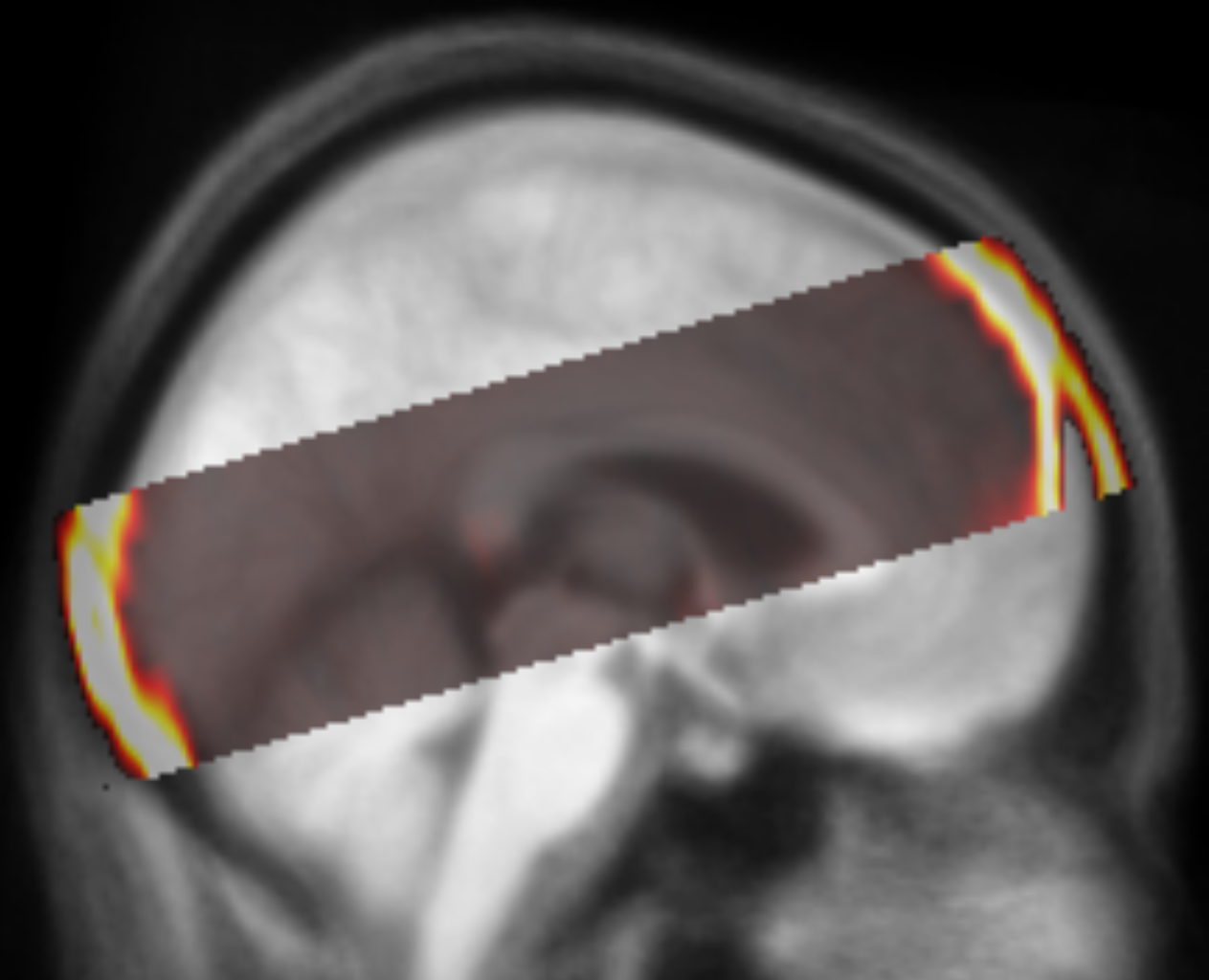}\label{subfigure:middle_group}
    \includegraphics[width=0.24\textwidth]{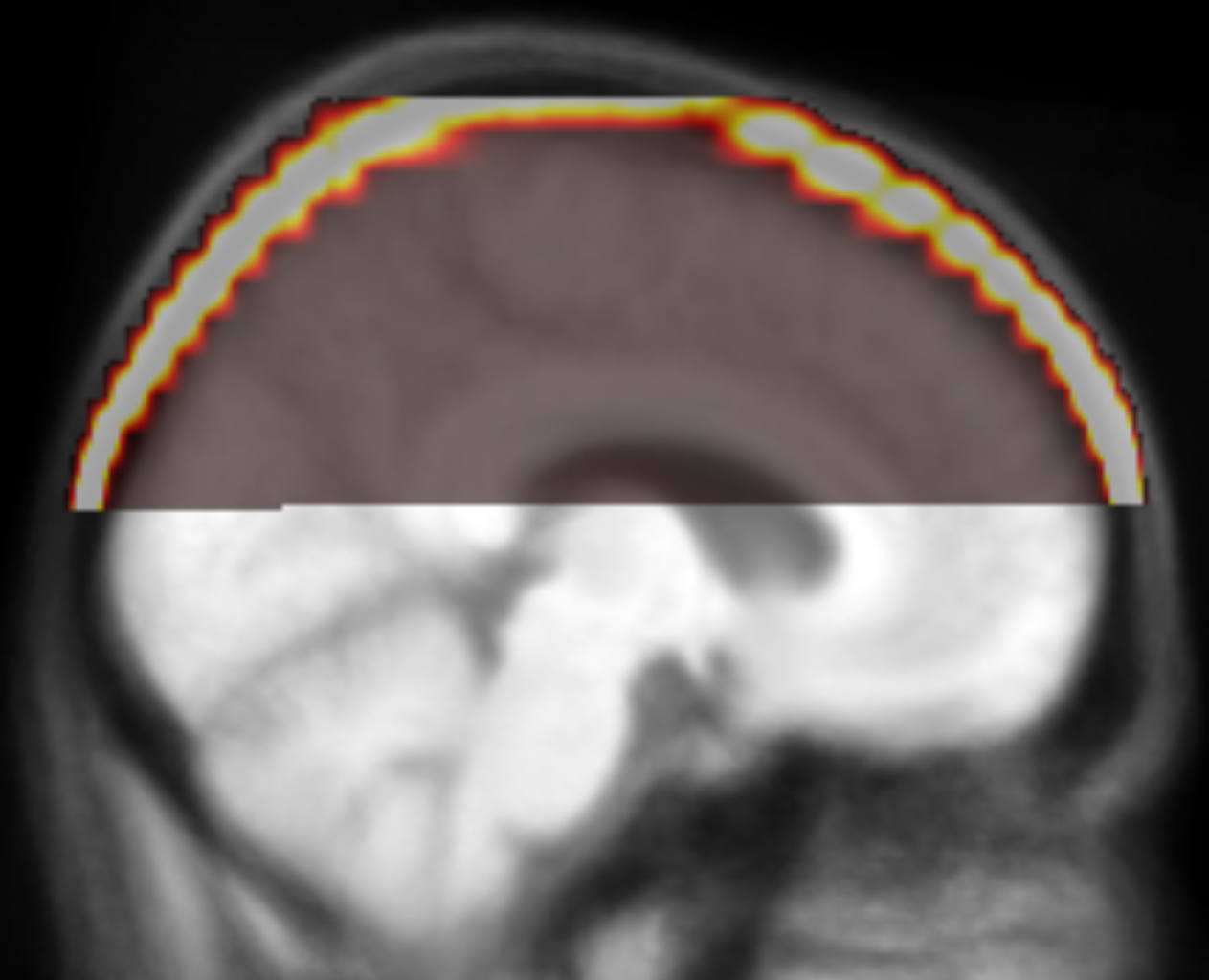}\label{subfigure:skull_vault_group}
    \caption{From left to right examples for the complete, skull base, medial, and skull vault subgroups. The series are overlayed on an \acrshort{mri} template from \cite{Farrell2009}.}
    \label{figure:completeness_groups}
\end{figure}

The \acrfull{ssim} \cite{Wang2004} is commonly used to compare image similarity. In contrast to more basic similarity methods, e.g. (inverse) mean squared error or cross-correlation, it normalises luminance and contrast. As a result, \acrshort{ssim} is a better fit for finding structural similarity in images with heterogeneous content \cite{Wang2009}. We compare \acrshort{ssim} scores of the registration results to the \acrshort{mri} templates (template from individuals aged 65-70 for younger group, template from ages 75-80 for older group) to find faulty registrations.

Comparing \acrshort{ssim} scores across CT series that cover different amounts and locations of the head is meaningless. Especially some older CT scanners produce separate series for the skull base and skull vault due to the need for different energies of radiation for each of these parts (modern scanners modulate energy delivery in real-time). We use the distribution of information presence over the z-axis in co-registered image sets to subgroup them into five categories: complete, skull base, medial, skull vault, and incomplete, before evaluating the similarity. We show exemplary cases of all subgroups except for the incomplete subgroup (as it is a collection of different types of series) in \autoref{figure:completeness_groups}.

\subsection{Superimposition quality control}

\begin{figure}
    \centering
    \includegraphics[width=0.2\textwidth]{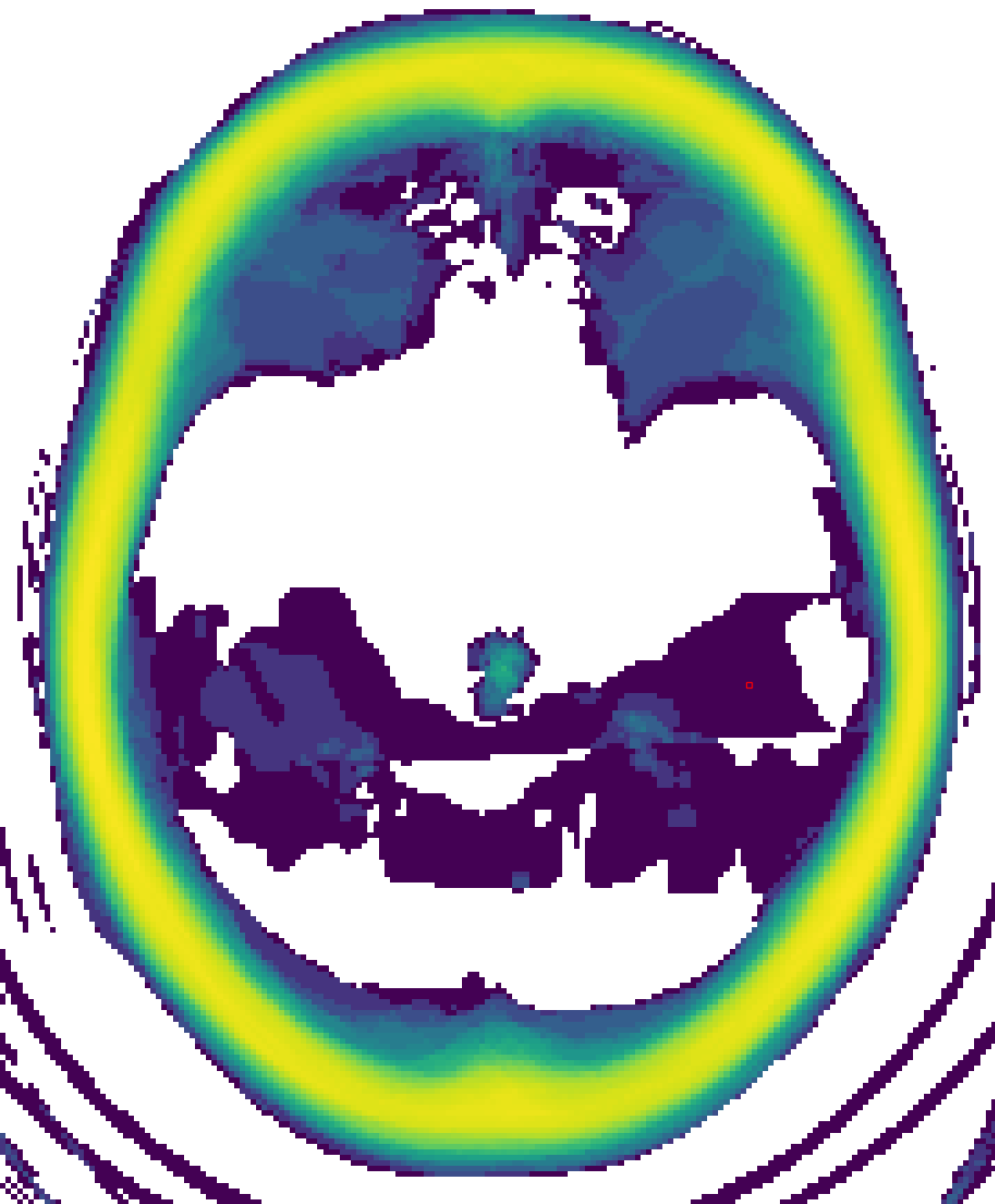}
    \includegraphics[width=0.2\textwidth]{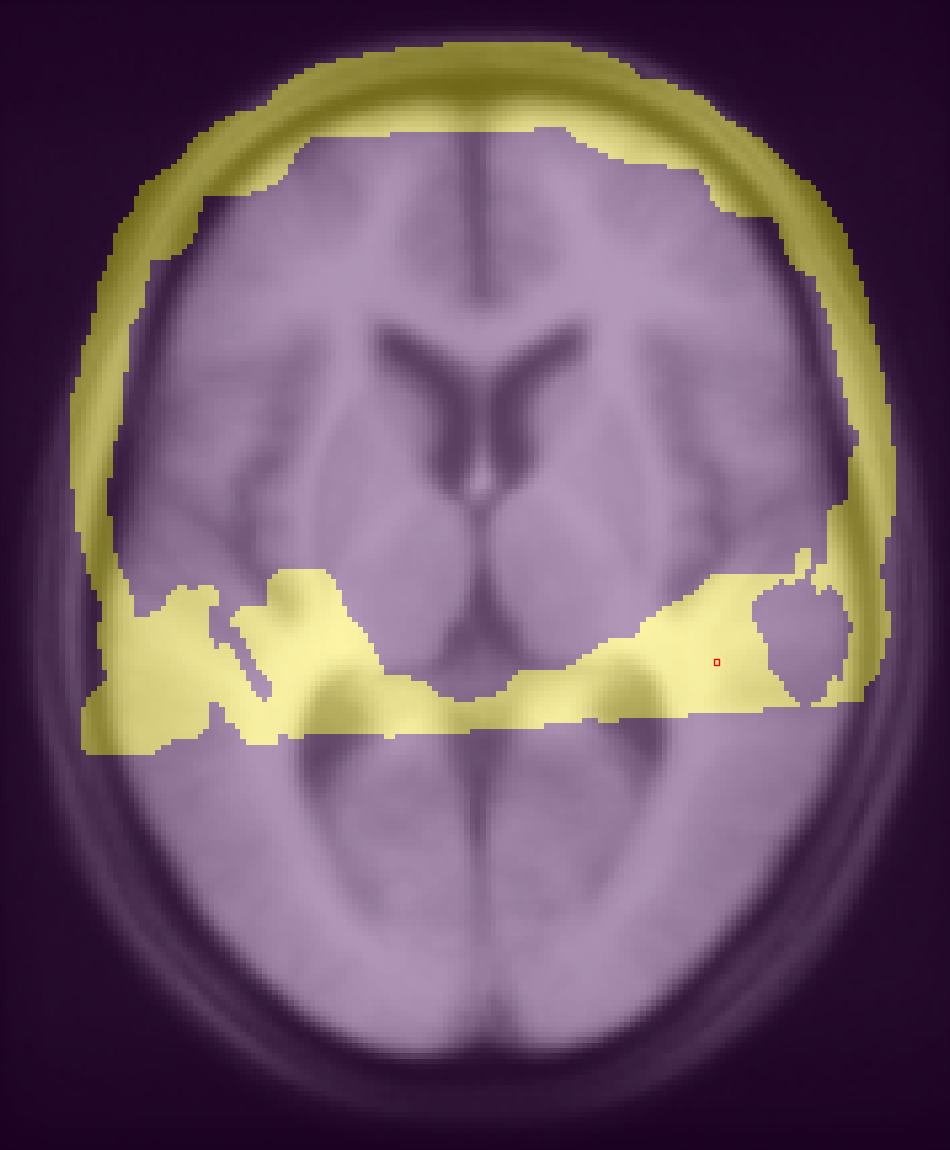}
    \includegraphics[width=0.2\textwidth]{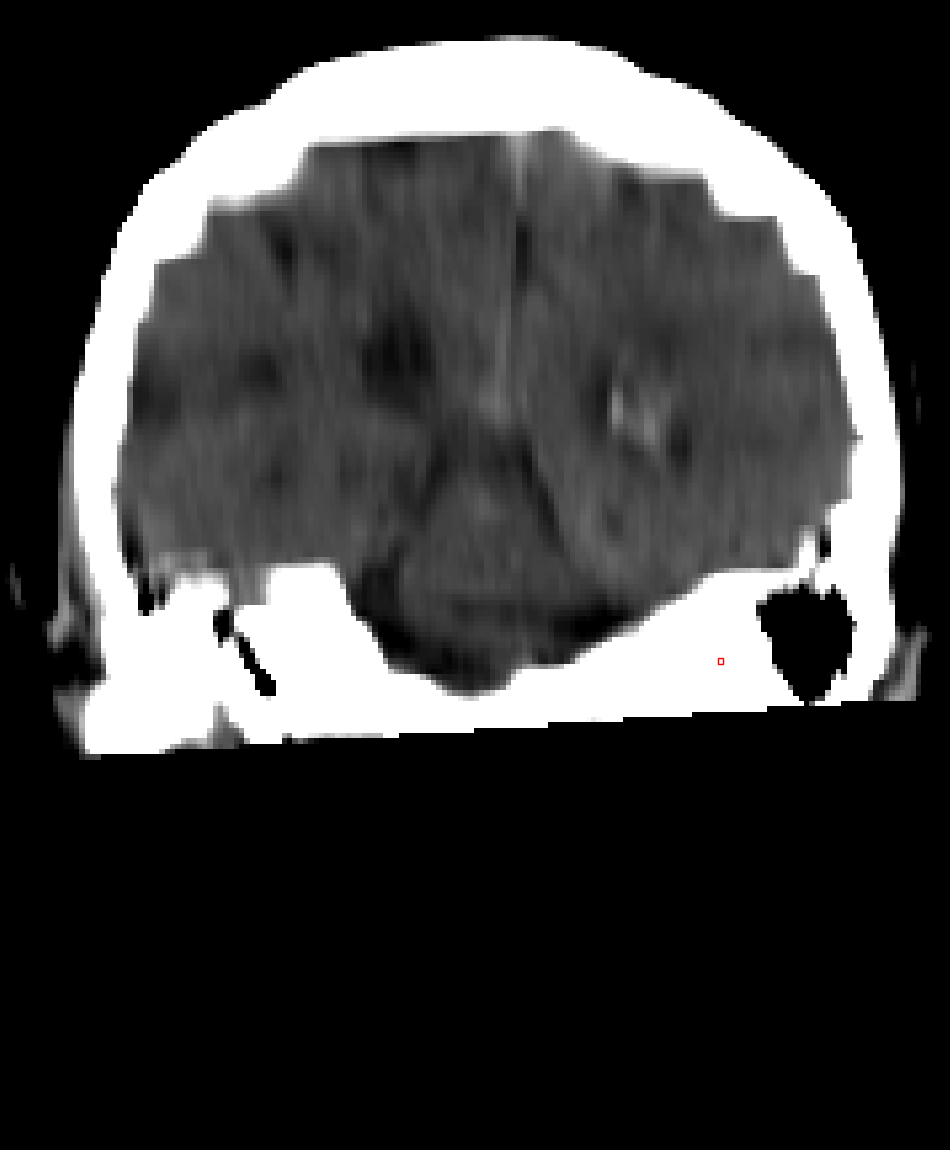}
    \caption{From left to right: 100 superimposed binary masks of co-registered series, a selected binary mask overlayed on an \acrshort{mri} template from \cite{Farrell2009}, the original registered series. Here, the coronal was erroneously registered to the axial plane.}
    \label{figure:superimpose_software}
\end{figure}

We built custom software to superimpose thresholded ($thresh = +100$ HU to include calcifications) binary masks derived from the co-registered CT scans to identify anomalies in the alignment with the template and high attenuation values at unexpected locations\footnote{https://github.com/bjin96/superimposition-tool}. The software provides three views (\autoref{figure:superimpose_software}). We use the first view to find anomalies in a large number of CT series. The superimposed image can be inspected in the axial plane and scrolled along the z-axis. Each of the voxels in the view can be selected if any of the binary masks has data at the specific location. If there are multiple binary masks at the selected location (i.e. multiple of the superimposed CT series have values above the threshold at the selected voxel), a single series can be chosen from a list. Upon selection, the alignment of the corresponding binary mask with the \acrshort{mri} template and the original registered series can be examined in the second and third views to inspect the nature of the anomaly and identify possible causes. Found anomalies are documented to a file with a textual comment. The software provides an efficient way to assess a large quantity of registered CT series and find cases with high attenuation values at anomalous locations indicating outliers and inaccurate registrations.

\section{Results}

\begin{figure}
    \centering
    \subfigure[]{\includegraphics[width=0.39\textwidth]{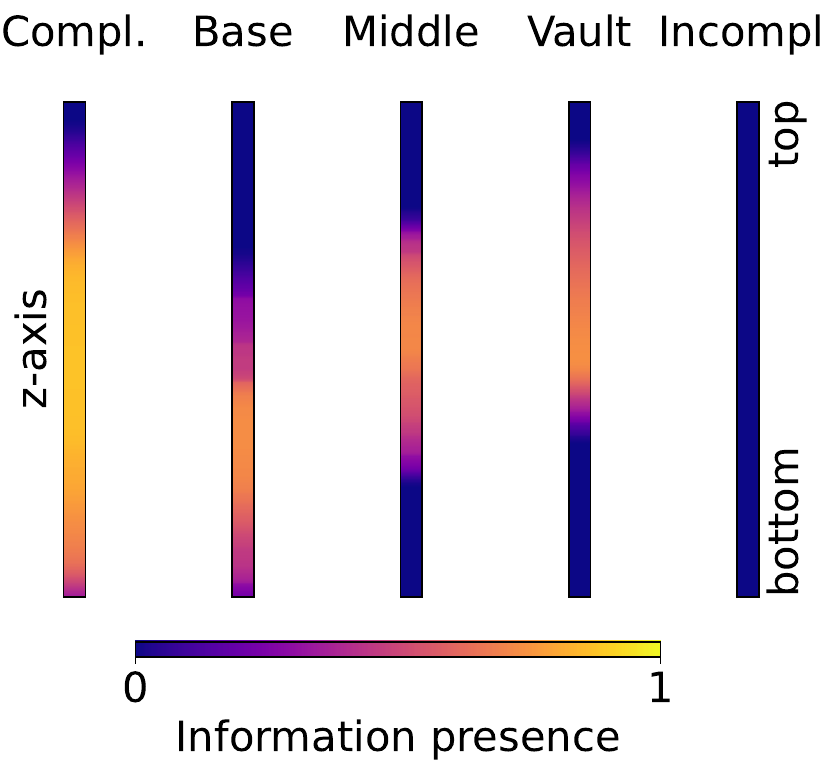}\label{subfig:group_information_presence}}
    \subfigure[]{\includegraphics[width=0.45\textwidth]{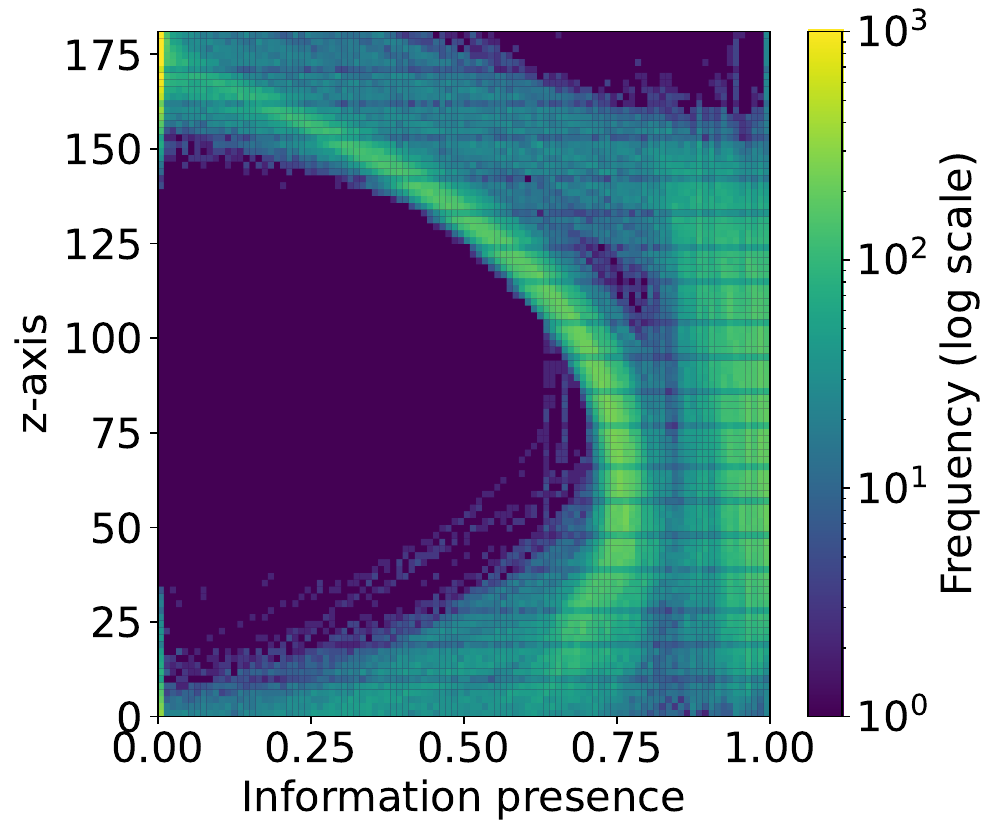}\label{subfig:information_presence_heatmap}}
    \caption{The information presence median distribution for the subgroups along the z-axis in (a) and a heat map of the information presence distribution for the complete subgroup in (b). Both figures are derived from the younger age group.} \label{figure:information_presence_distribution}
\end{figure}

We present the results from our proposed quality control methodology and our overall pipeline results. We summarise the information presence distribution of each subgroup with its median along the z-axis in Figure \ref{subfig:group_information_presence}, visualising the effectiveness of our proposed metric in classifying the series based on completeness. In Figure \ref{subfig:information_presence_heatmap} we depict the information presence distribution along the z-axis for the complete subgroup as a heat map (more in supplementary Figure 4), where the colouring denotes the number of series with the specific information presence at the specific slice (in log scale to make less frequent anomalies visible). We observe a wider spread of information presence in the lower slices (i.e., low values in the z-axis assuming standard orientation/positioning) due to differences in the starting position of the scanning and gantry tilt correction. While most series follow a hyperbolic curve from an information presence of about 0.6, peaking at about 0.75 and ending at 0, a larger group of series show higher information presence. These series exhibit relatively higher attenuation values inside compared to outside the scanner's field of view, which can be masked by appropriate windowing.

\begin{figure}
    \centering
    \includegraphics[width=0.96\textwidth]{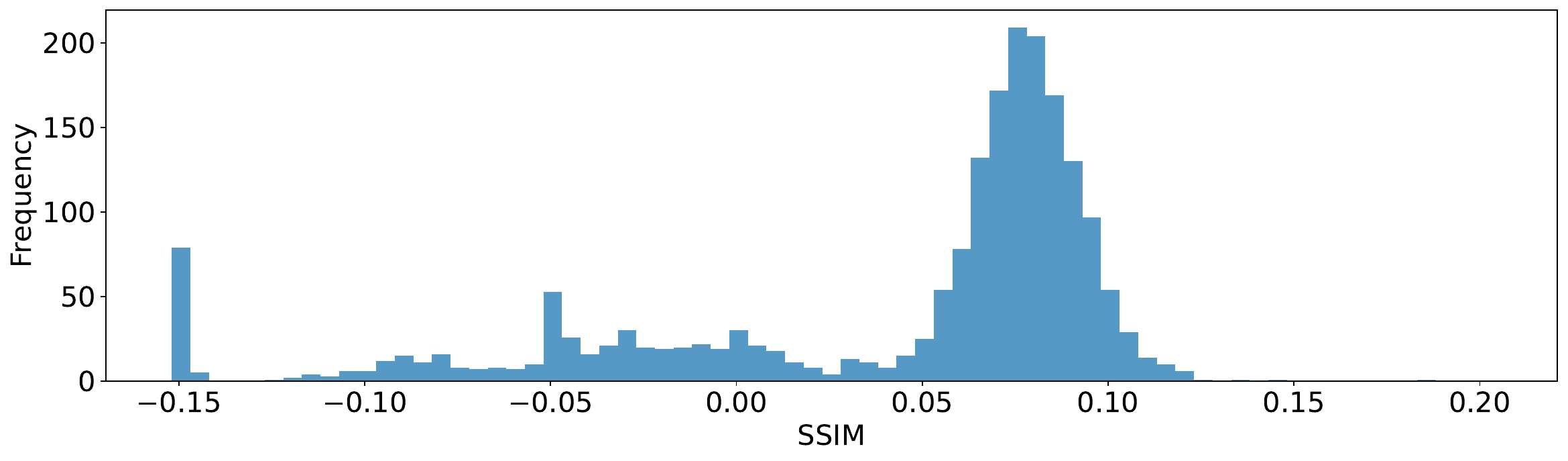}
    \includegraphics[width=0.47\textwidth]{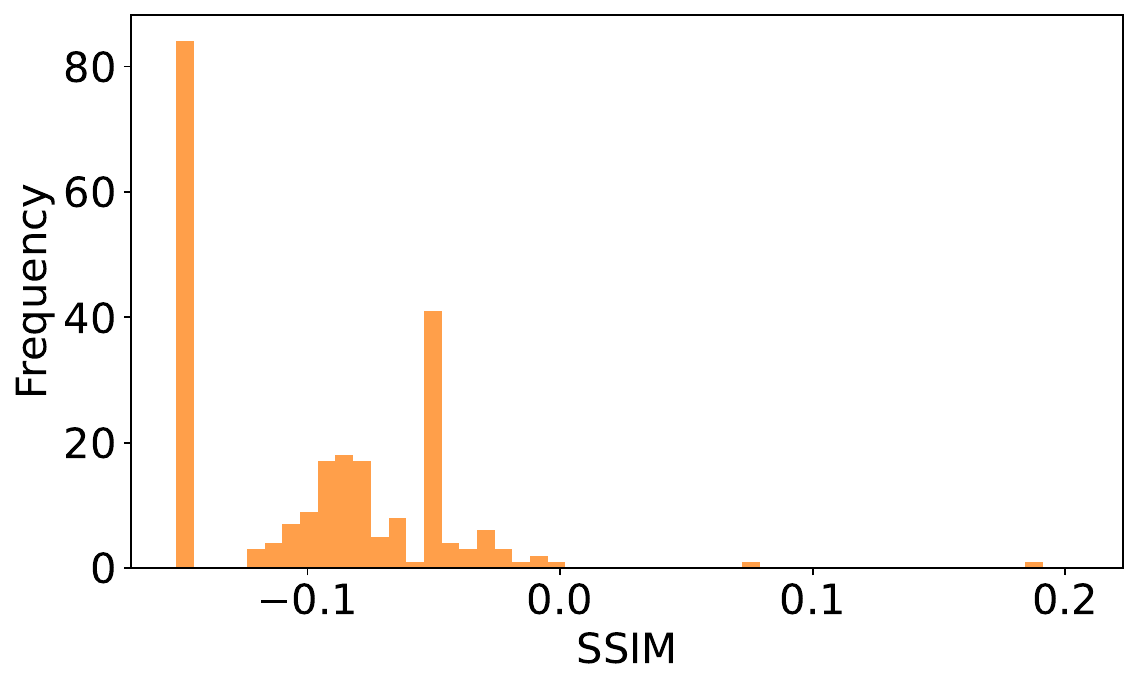}
    \includegraphics[width=0.47\textwidth]{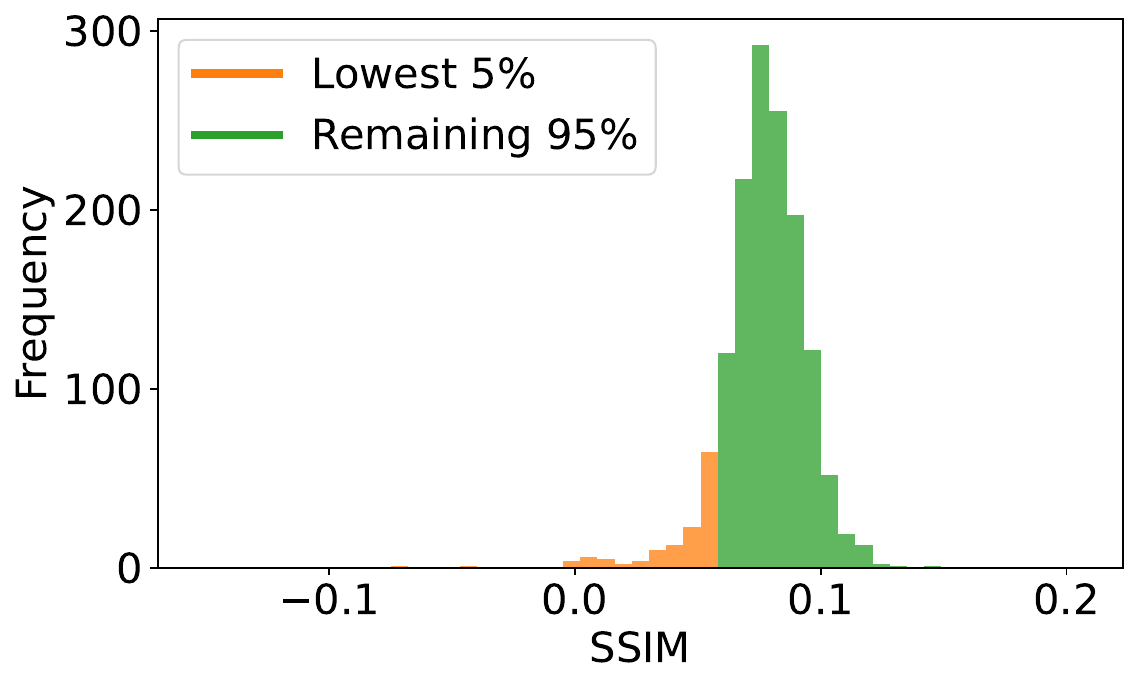}
    \caption{SSIM score distributions for the entire younger age group at the top and the incomplete (left) and complete (right) subgroups below. Orange bars indicate visually inspected series and green automatic acceptance.}
    \label{figure:ssim_distribution}
\end{figure}

For \acrshort{ssim} quality control, we show plots of the \acrshort{ssim} distribution for the younger age group in \autoref{figure:ssim_distribution}. As the two extremes, we highlight the incomplete and complete subgroup (more in supplementary Figure 5). The incomplete subgroup predominantly comprises \acrshort{ssim} scores below zero, indicating a negative correlation between the series and the registered template, the complete subgroup consists of mostly positive \acrshort{ssim} scores. Hence, we manually inspect (orange) all the incomplete series and the series at the lower end of \acrshort{ssim} scores for the complete subgroup. Based on the inspection of the lowest 5\% of \acrshort{ssim} scores for the complete, skull base, and skull vault subgroup and the entire medial and incomplete subgroups, we discard 1,040 series. Since there is redundancy in the series of a CT scan (e.g. soft tissue and bone kernel or thick and thin slice series of the same scan), we only lose all scan data from 24 patients at this step.

We inspect the entire remaining dataset (7,493 series) using our superimposition quality control with the effort of looking at 75 series by assessing superimposed image sets that summarise 100 series. We find inaccurate registrations not flagged by low similarity scores and additional anomalous data that affect the registration accuracy and subsequent \acrshort{roi} quality. Examples include series with missing slice data in the middle of the image series, artefacts outside and inside of the patient's head, information leak between slices, incorrect slice order, patient movement, registrations in the wrong view (i.e. coronal or sagittal plane is registered to the axial plane), and inaccurate registrations due to remaining gantry tilt. We only exclude cases where our \acrshortpl{roi} are not captured by the registered series.

An overview of our pipeline steps with the number of remaining series and unique patients at the end of each step is given in \autoref{table:pipeline_overview}. In total, we reject 4,326/10,659 series (41\%) and the quality control accounts for 1,450 series (34\%) of the rejections, leaving 6,333 series as the output of the pipeline.

\begin{table}
    \centering
    \caption{Summary of the pipeline steps and data loss after each step.}
    \label{table:pipeline_overview}
    \begin{tabular}{|l|r|r||r|r|}
    \hline
    \textbf{Step} &  \textbf{No. of series} & \textbf{Change} & \textbf{No. of patients} & \textbf{Change}\\
    \hline
    Source CT series & 10,659 & 0 & 2,578 & 0\\
    Conversion to NIfTI & 10,638 & -21 & 2,577 & -1\\
    Limit to axial series & 9,000 & -1,638 & 2,570 & -7\\
    Remove localiser series & 8,558 & -442 & 2,570 & 0\\
    Affine co-registration & 8,533 & -25 & 2,570& 0\\
    \hline
    Similarity QC & 7,493 & -1,040 & 2,546 & -24\\
    Superimposition QC & 7,083 & -410 & 2,489 & -57\\
    \hline
    Containing $\geq$1 ROI & 6,337 & -746 & 2,351 & -138\\
    \hline
    \textbf{Total change} & \textbf{-41\%} & \textbf{-4,322} & \textbf{-9\%} & \textbf{-227}\\
    \hline
    \hline
    \end{tabular}
\end{table}

\section{Discussion and Conclusion}

General medical imaging pipelines can handle large-scale datasets and help gaining an overview of the data \cite{Denner2023}. More targeted, Muschelli recommends pre-processing steps for \acrshort{ct} head scans \cite{Muschelli2019}. Closer to our work and purpose, Bortsova et al. developed a pipeline to automatically pre-process their scans for \acrshort{iac} segmentation with deep neural networks \cite{Bortsova2021}, but they omit their quality control process. Given the research nature of their data, extensive quality control is presumably unnecessary. Quality control methods exist, for instance with the tool MRIQC \cite{Esteban2017} developed for the \acrshort{mri} domain, but we focus on ensuring registration quality. For heterogeneous real-world clinical data, we find that straightforward co-registration of the \acrshort{ct} series to templates with pre-defined \acrshortpl{roi} leads to a considerable number of inaccurate registrations which would limit the performance of a deep learning model trained on the resulting dataset. Our pipeline, which includes highly efficient quality control methods, enables checking large number of CT scans and their respective registrations for anomalies to output high-quality \acrshortpl{roi} around major intracranial arteries.

Fontanella et al. developed a pipeline on the same \acrshort{ist3} dataset tailored towards analysis in brain soft tissue \cite{Fontanella2023}. Our pipeline leverages on their work, with adaptions to include CT series that otherwise would not be considered, substantially increasing the dataset size. Most importantly, we propose additional quality control methods and evaluations. Fontanella et al. use principal component analysis and clustering of the transformation matrices to detect incorrect registrations, hence they rely on transformation information to determine registration quality. Our methods identify more general incorrect registrations by calculating similarity in comparable groups and with a superimposition tool leveraging the CT image series contents beyond transformation information.

As clinical data can exhibit a list of expected and unexpected anomalies, we developed our pipeline to include minimal human intervention to support a thorough understanding of the data and instil confidence in the resulting dataset. Our results will facilitate later efforts at automating additional pipeline components with our description of different data anomalies. Ultimately, we will deploy the pipeline including trained deep learning segmentation models on a population-scale dataset of routinely collected clinical imaging to link \acrshort{iac} with clinical outcomes of neurovascular diseases such as stroke. Since our pipeline rejects a considerable number of series of confirmed inaccurate registrations, we suggest further research into designing and comparing end-to-end deep learning methods for \acrshort{iac} segmentation, which omit the intermediate co-registration. Such approaches will make model training more challenging and computationally expensive, but potentially more inclusive of heterogeneous routine clinical data.

\begin{credits}

\subsubsection{\ackname}
We thank the IST-3 collaborative group for the generous provision of data from the \acrshort{ist3} and all patients who participated in the study. \acrshort{ist3} was funded from many sources but chiefly the UK Medical Research Council (MRC G0400069 and EME09-800-15) and the UK Stroke Association. The \acrshort{ist3} data is managed and stored within the Systematic Management, Archiving and Reviewing of Trial Images Service (SMARTIS) at the University of Edinburgh. 
B. J. is funded by the Medical Research Council [grant number MR/W006804/1]. M.V.H. is funded by The Row Fogo Charitable Trust [grant number BROD.FID3668413].
G.M. is the Stroke Association Edith Murphy Foundation Senior Clinical Lecturer (SA L-SMP 18\textbackslash1000).

\subsubsection{\discintname}
The authors have no competing interests to declare that are relevant to the content of this article. 

\end{credits}

%
%
%
\bibliographystyle{splncs04}
\bibliography{samplepaper}

\end{document}


\section*{Pre-processing and quality control of large clinical CT head datasets for intracranial arterial calcification segmentation -- supplementary materials}

\begin{figure}
    \centering
    \includegraphics[width=0.95\textwidth]{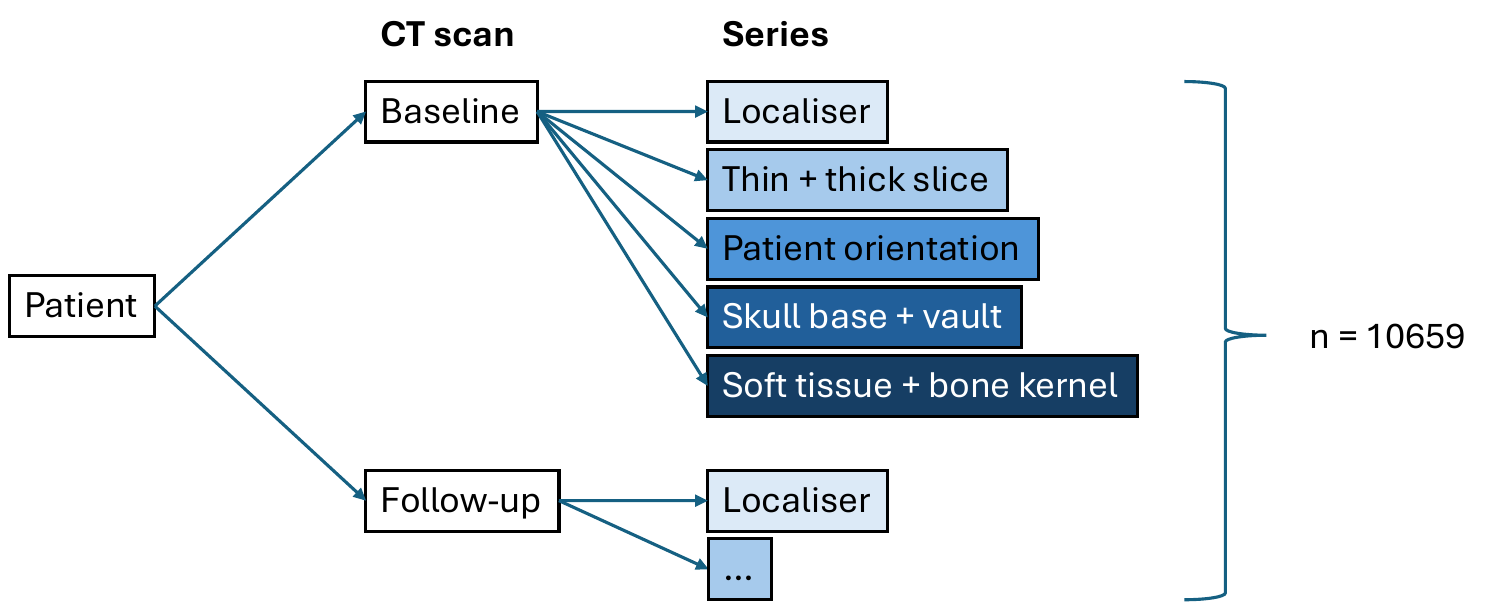}
    \caption{Example of the data structure and the types of series found in a CT scan for a patient in the \acrshort{ist3} dataset.}
    \label{supp:study_overview}
\end{figure}
\begin{figure}[hbt!]
    \centering
    \includegraphics[width=0.135\textwidth]{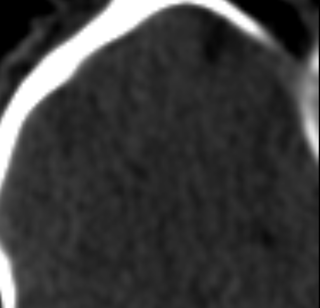}
    \includegraphics[width=0.135\textwidth]{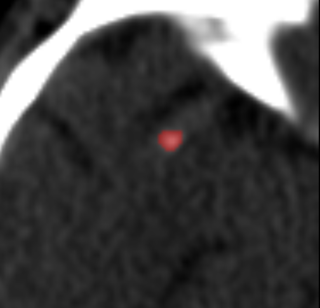}
    \includegraphics[width=0.135\textwidth]{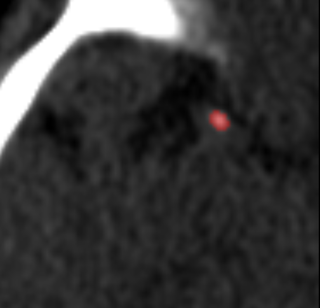}
    \includegraphics[width=0.135\textwidth]{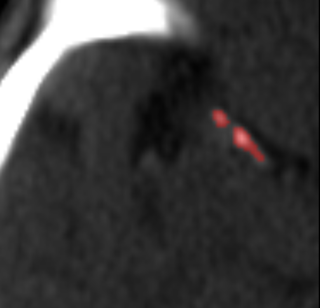}
    \includegraphics[width=0.135\textwidth]{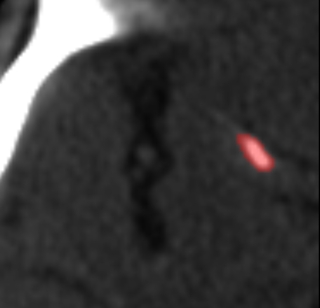}
    \includegraphics[width=0.135\textwidth]{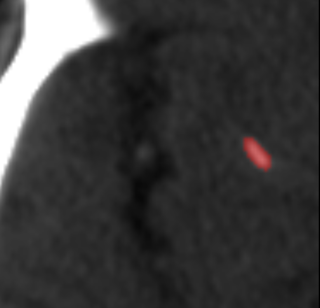}
    \includegraphics[width=0.135\textwidth]{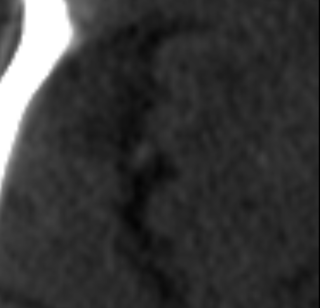}
    
    \includegraphics[width=0.135\textwidth]{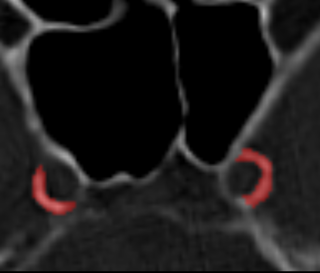}
    \includegraphics[width=0.135\textwidth]{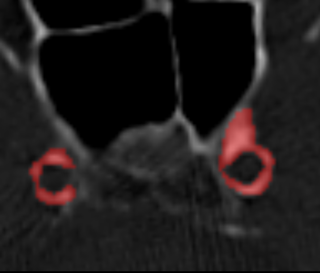}
    \includegraphics[width=0.135\textwidth]{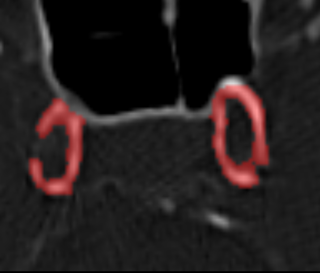}
    \includegraphics[width=0.135\textwidth]{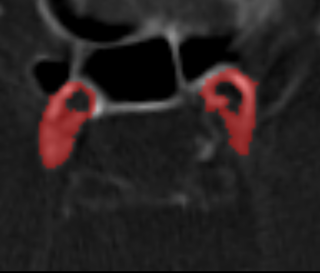}
    \includegraphics[width=0.135\textwidth]{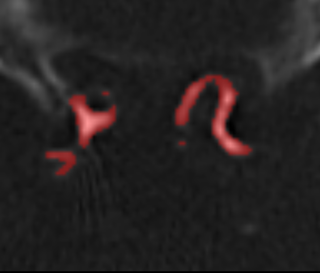}
    \includegraphics[width=0.135\textwidth]{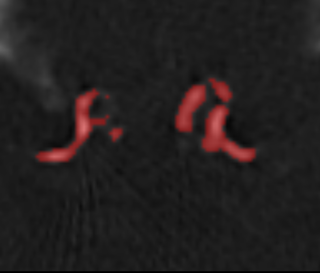}
    \includegraphics[width=0.135\textwidth]{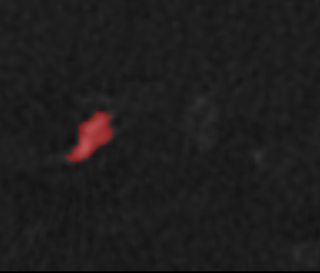}
    
    \includegraphics[width=0.135\textwidth]{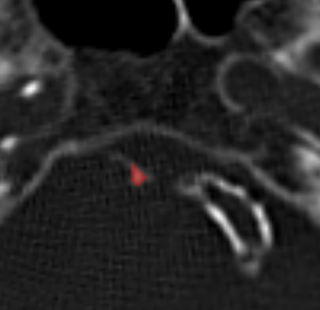}
    \includegraphics[width=0.135\textwidth]{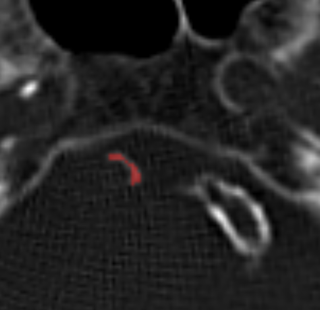}
    \includegraphics[width=0.135\textwidth]{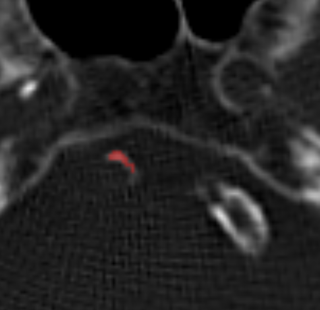}
    \includegraphics[width=0.135\textwidth]{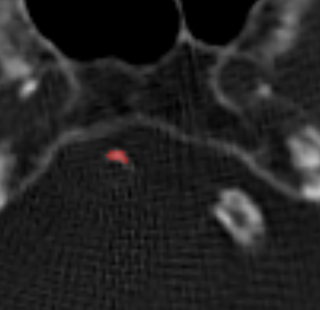}
    \includegraphics[width=0.135\textwidth]{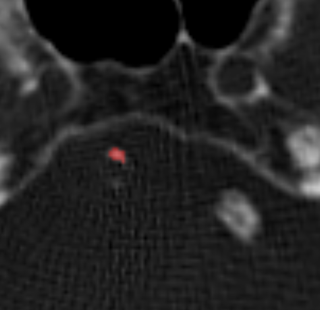}
    \includegraphics[width=0.135\textwidth]{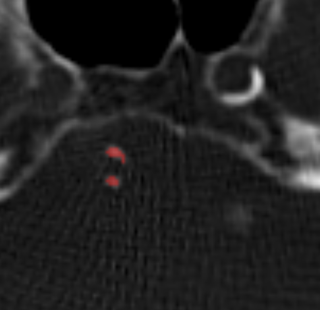}
    \includegraphics[width=0.135\textwidth]{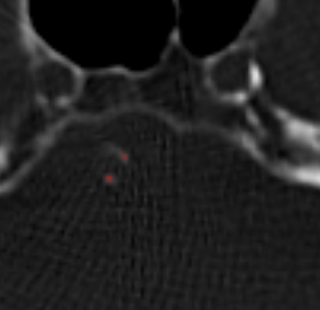}
    
    \includegraphics[width=0.135\textwidth]{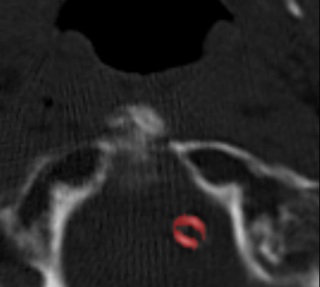}
    \includegraphics[width=0.135\textwidth]{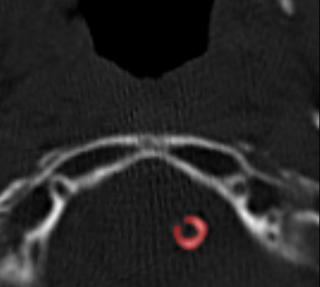}
    \includegraphics[width=0.135\textwidth]{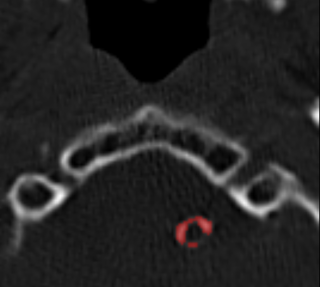}
    \includegraphics[width=0.135\textwidth]{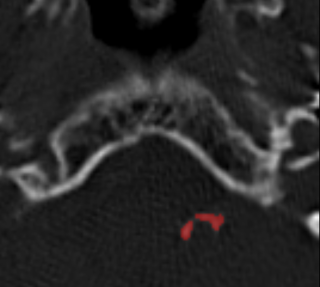}
    \includegraphics[width=0.135\textwidth]{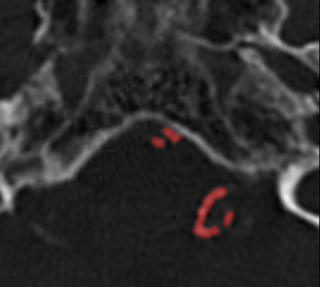}
    \includegraphics[width=0.135\textwidth]{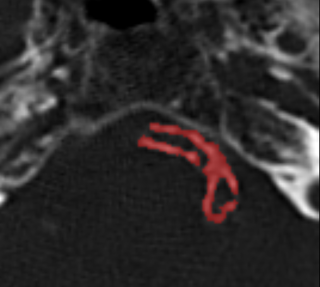}
    \includegraphics[width=0.135\textwidth]{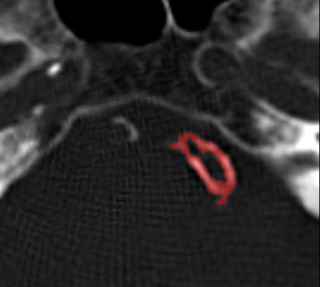}
    
    \caption{\acrshortpl{roi} from example CT series with arterial calcifications highlighted (red). From top to bottom, M1 segment of a right middle cerebral artery, cavernous segment of an internal carotid artery, a basilar artery, and a vertebral artery.}
    \label{suppl:example_roi_calcifications}
\end{figure}
\begin{figure}[hbt!]
    \centering
    \includegraphics[width=0.118\textwidth]{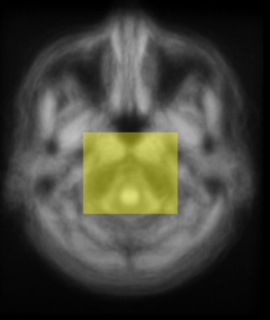}
    \includegraphics[width=0.118\textwidth]{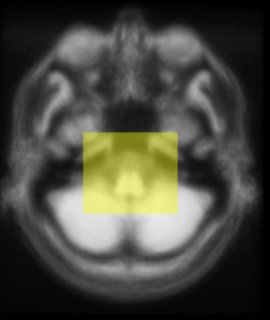}
    \includegraphics[width=0.118\textwidth]{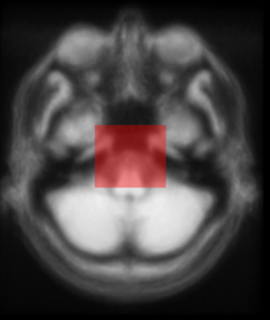}
    \includegraphics[width=0.118\textwidth]{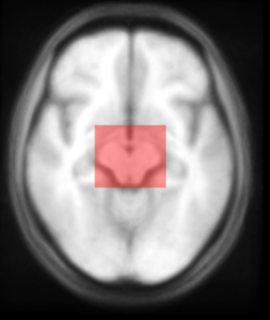}
    \includegraphics[width=0.118\textwidth]{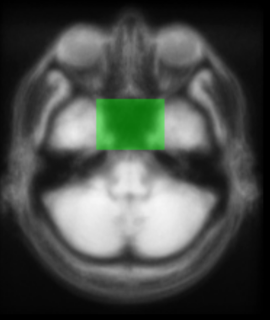}
    \includegraphics[width=0.118\textwidth]{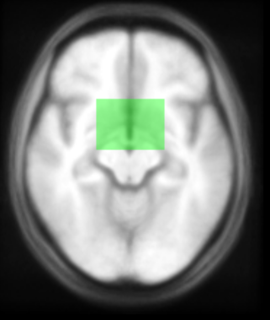}
    \includegraphics[width=0.118\textwidth]{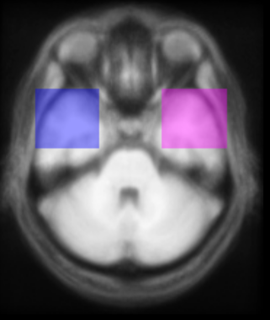}
    \includegraphics[width=0.118\textwidth]{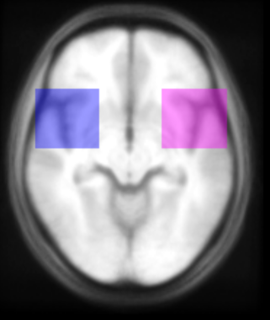}
    \caption{\acrshortpl{roi} defined by their inferior (left) and superior (right) slices on an \acrshort{mri} template. Vertebral (yellow), basilar (red), cavernous segment of the internal carotid (green), left (purple) and right (blue) M1 segment of the middle cerebral arteries.}
    \label{suppl:template_rois}
\end{figure}

\begin{figure}
    \centering
    \includegraphics[width=0.45\textwidth]{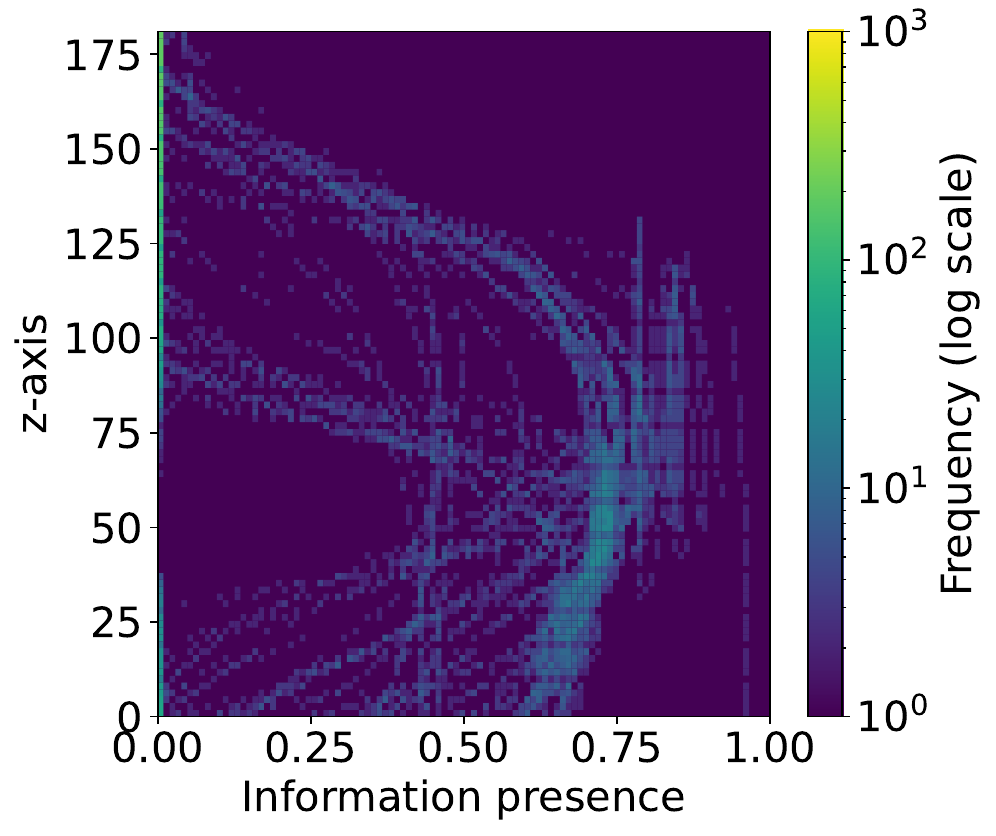}
    \includegraphics[width=0.45\textwidth]{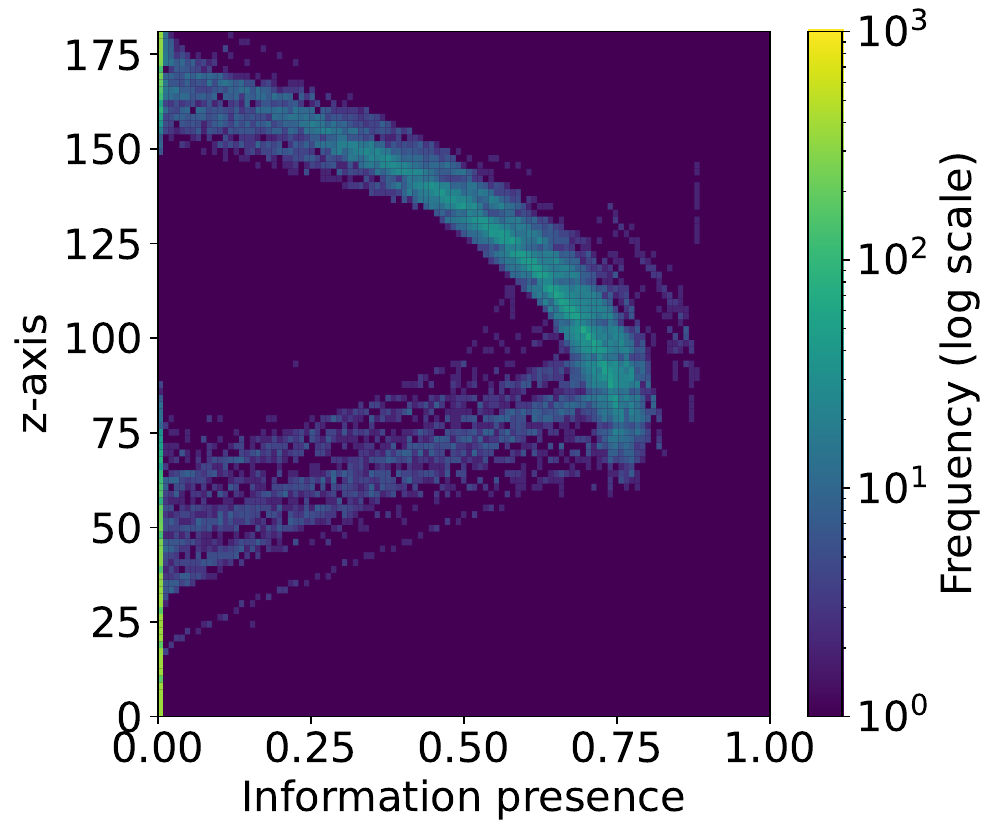}
    \caption{Heat maps of information presence distribution for series registered to the younger age MRI template. Skull base (left) and skull vault (right) subgroups.}
    \label{suppl:young_heatmaps}
\end{figure}

\begin{figure}[hbt!]
    \centering
    \includegraphics[width=0.49\textwidth]{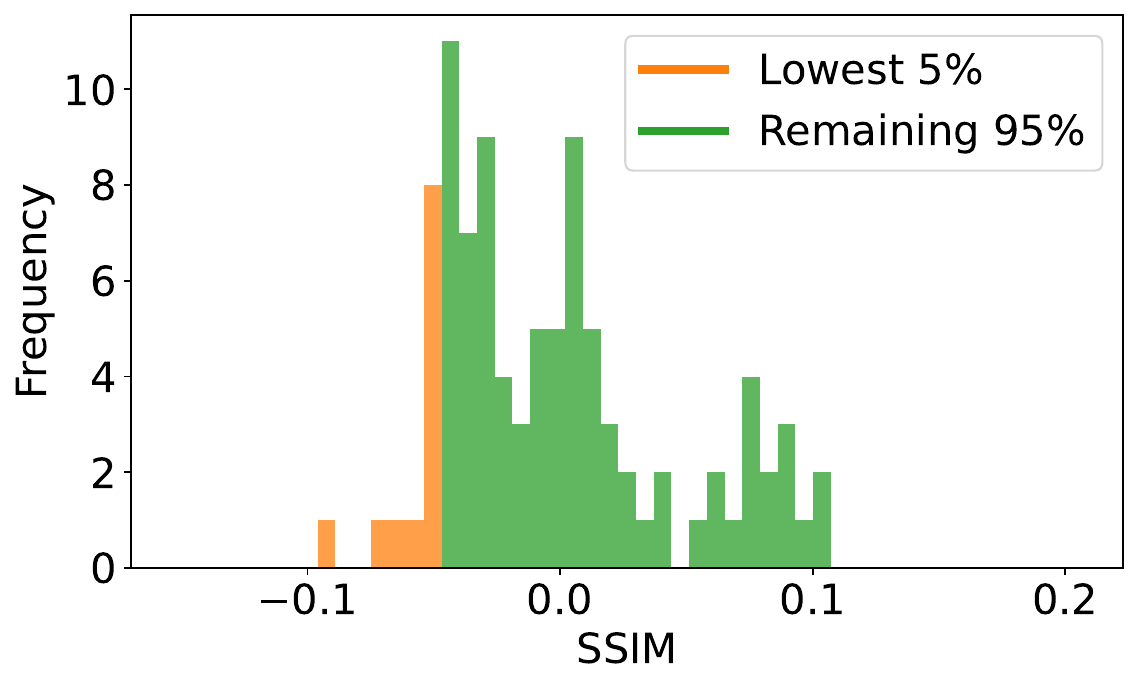}
    \includegraphics[width=0.49\textwidth]{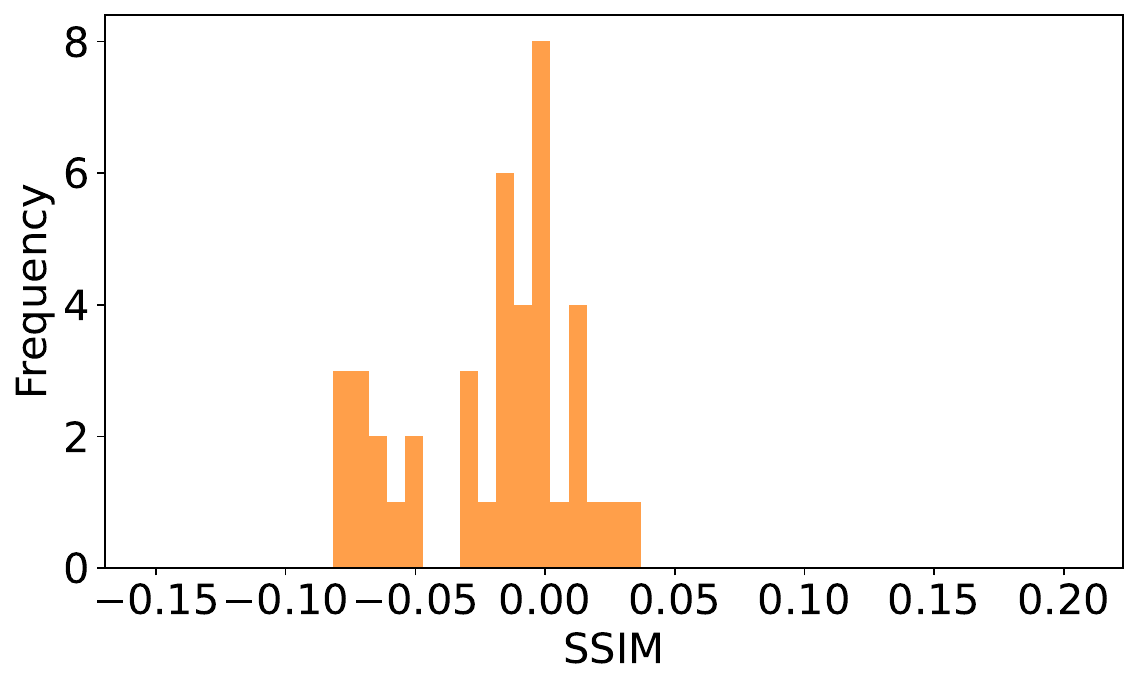}
    \includegraphics[width=0.49\textwidth]{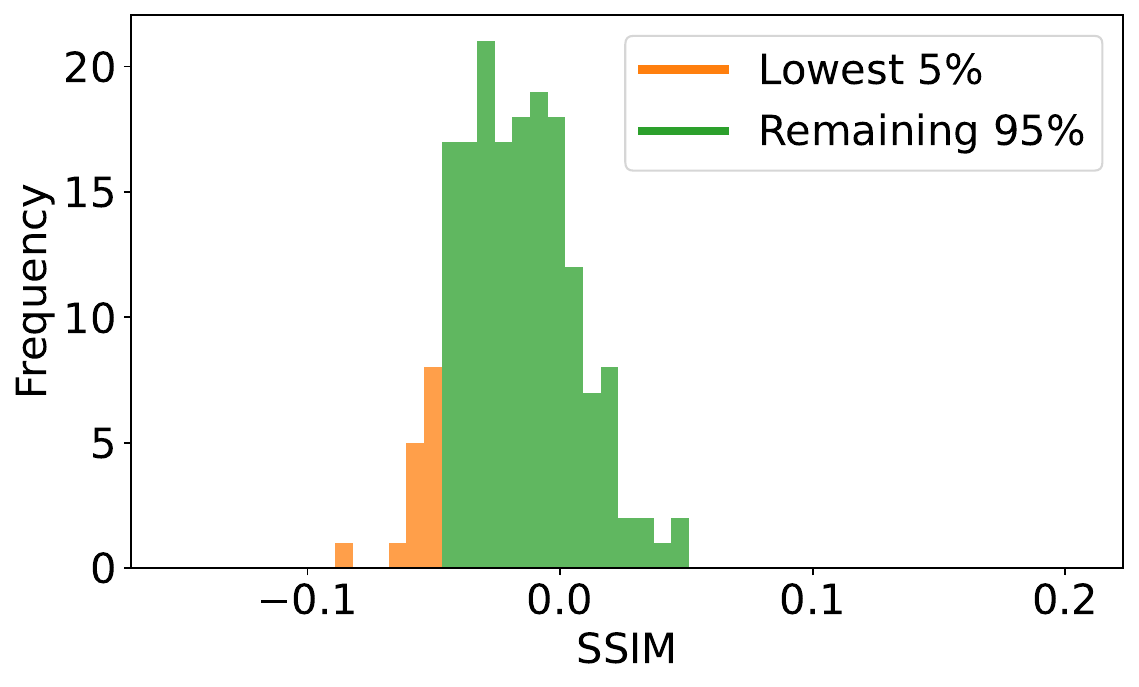}
    \caption{SSIM score distributions for the series registered to the younger age MRI template. Left-right, top-bottom: skull base, medial, and skull vault subgroups. Orange bars indicate visually inspected series, green automatic acceptance.}
    \label{suppl:young_ssim_distribution}
\end{figure}